\documentclass[twocolumn,pra,superscriptaddress,showpacs]{revtex4}
\usepackage{graphics}

\begin{document}
\title{Thermal and back-action noises in dual-sphere gravitational-waves detectors}
\author{T. Briant}
\affiliation{Laboratoire Kastler Brossel, CNRS, Ecole Normale
Sup\'{e}rieure and Universit\'{e} P. et M. Curie, Case 74, 4 place
Jussieu, F75252 Paris Cedex 05, France}
\author{M. Cerdonio}
\affiliation{INFN Padova Section and Department of Physics,
University of Padova, via Marzolo 8, I-35100 Padova, Italy}
\author{L. Conti}
\affiliation{INFN Padova Section and Department of Physics,
University of Padova, via Marzolo 8, I-35100 Padova, Italy}
\author{A. Heidmann}
\affiliation{Laboratoire Kastler Brossel, CNRS, Ecole Normale
Sup\'{e}rieure and Universit\'{e} P. et M. Curie, Case 74, 4 place
Jussieu, F75252 Paris Cedex 05, France}
\author{A. Lobo}
\affiliation{Departament de Fisica Fonamental, Universitat de
Barcelona, Diagonal 647, E-08028 Barcelona, Spain}
\author{M. Pinard}
\affiliation{Laboratoire Kastler Brossel, CNRS, Ecole Normale
Sup\'{e}rieure and Universit\'{e} P. et M. Curie, Case 74, 4 place
Jussieu, F75252 Paris Cedex 05, France}
\date{December 19, 2002}

\begin{abstract}
We study the sensitivity limits of a broadband gravitational-waves
detector based on dual resonators such as nested spheres. We
determine both the thermal and back-action noises when the
resonators displacements are read-out with an optomechanical
sensor. We analyze the contributions of all mechanical modes,
using a new method to deal with the force-displacement transfer
functions in the intermediate frequency domain between the two
gravitational-waves sensitive modes associated with each
resonator. This method gives an accurate estimate of the
mechanical response, together with an evaluation of the estimate
error. We show that very high sensitivities can be reached on a
wide frequency band for realistic parameters in the case of a
dual-sphere detector.
\end{abstract}


\maketitle

\section{Introduction}

Dual systems, in the form of nested spheres
\cite{Cerdonio01,Conti02a} and nested cylinders \cite{Bonaldi02}
have been recently proposed as wideband gravitational-waves
detectors, of spectral sensitivity complementary in frequency to
advanced interferometric detectors. Dual spheres are based on two
spherical masses nested together: the inner mass is a full sphere
while the other one is a hollow sphere. The radii match so that
only a small gap separates the two bodies.

Fundamental modes of both spheres are quadrupole modes sensitive
to gravitational waves. By a proper choice of the mechanical and
geometrical characteristics of the two spheres, the fundamental
mode of the inner sphere occurs at a frequency 2 or 3 times larger
than the fundamental mode of the hollow one, without any sensitive
modes in the intermediate frequency domain between these two
fundamental modes. This frequency domain is of particular interest
as the spheres displacements caused by a gravitational wave are
out of phase by $\pi$ radians, thus leading to a measurable
variation of the gap between the spheres.

To get a wideband detector, very sensitive and non-resonant
displacements sensors are needed. An efficient technique consists
in using optomechanical sensors based on a high-finesse
Fabry-Perot cavity \cite{Hadjar99,Cohadon99,Conti02b}. One mirror
of the cavity is coated on the inner side of the hollow sphere,
whereas the other mirror is coated on the solid sphere.
Measurement of the phase of the field reflected by such a linear
cavity thus provides information on the gap variation between the
two spheres, at the radial position of the sensor. One can choose
a strategy to set the number of optomechanical sensors and their
location in such a way to reconstruct the sphere motion
\cite{Merkovitz97,Lobo00}.

A key point for broadband operation is the sensitivity of the
optomechanical sensor which actually depends on its coupling with
the mechanical modes of the system. Since the frequencies of
interest are between two mechanical resonances, dual systems are
conceptually different from bars and interferometers. With bars
one has to deal with a single resonator, driven by the
gravitational wave at its fundamental frequency. The treatment of
the bar excitation by the gravitational wave, by the thermal noise
in the bar and in the transducer, and by the back-action noise
induced by the sensor is relatively straightforward, including the
understanding of the Standard Quantum Limit. One gets a prediction
of the narrow-band spectral sensitivity as a function of relevant
parameters, dealing basically with few isolated modes
\cite{Price87}. For the thermal noise, correlations may arise
between the two modes originating from the tight coupling between
the bar and the resonant transducer, when their mechanical quality
factors are very different \cite{Majorana97,Conti02b}, but the
noise spectral behavior does not suffer dramatic changes.

In the case of interferometers the gravitational wave drives a set
of masses, which can be considered free above the pendulum
resonant frequencies of the suspensions. Again, the sources of
noise can be spelled out, down to the quantum limit and the
wideband spectral sensitivity can be predicted as a function of
relevant parameters \cite{Hello98}.

Dual systems are conceptually different in that one has to deal
neither at resonance as with bars, nor far from resonant modes as
with interferometers, but rather in between resonant modes. The
difficulty is then to write a mechanical transfer function for the
system valid in this unusual frequency range.

In this paper we determine the limits induced both by the thermal
noise of the spheres and by the quantum fluctuations of light,
including the measurement noise and the back-action effects of
light on the dual system. We show in the case of a single sensor
that the limit of sensitivity can be expressed in terms of
mechanical transfer functions characterizing the optomechanical
coupling of light with the two spheres.

We illustrate this behavior in the case where only the excitation
of fundamental modes of each sphere is taken into account, showing
a new effect of back-action noise cancellation in the frequency
domain between the two modes. This cancellation results from a
destructive interference between radiation pressure effects on
both spheres and leads to an increased sensitivity in between the
two resonances.

We then develop a simple approach to take into account all other
mechanical modes of the spheres. This method allows to estimate
the mechanical transfer functions in the intermediate frequency
range between the resonances and gives an upper bound for the
estimate error. As a result we determine the sensitivity of a
dual-sphere detector and show that a spectral strain sensitivity
better than $10^{-22}$ Hz$^{-1/2}$ can be obtained over a wide
frequency range from $1$ kHz to $3.5$ kHz.

In Section \ref{Optomechanical_Sensor} we present the basic
principles of the gravitational-waves detection by a dual system
with a transducer based on an optomechanical sensor. Section
\ref{Noises} is devoted to the determination of noises,
illustrated in the case of a dual system with only two mechanical
resonances. In section \ref{Mechanical_Susceptibility} we derive
the mechanical transfer functions, taking into account all
mechanical modes. Results for a Beryllium dual-sphere detector are
presented in Section \ref{Results}.

\section{Dual detector with optomechanical sensor}
\label{Optomechanical_Sensor}

\begin{figure}
{\resizebox{6 cm}{!}{\includegraphics{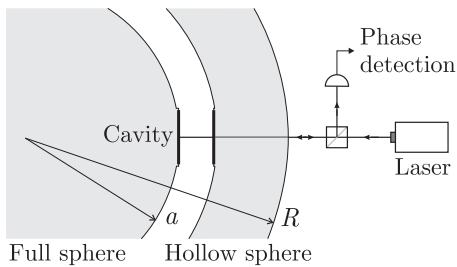}}}
\caption{\label{Fig_Setup}Scheme of the system studied in the
paper. The small gap between the two spheres of the
gravitational-waves detector (inner radius $a$, external radius
$R$) is measured by an optomechanical sensor made of a
high-finesse single-ended cavity. The phase of the reflected field
is sensitive to the differential displacement of the mirrors.}
\end{figure}

The scheme of the dual-sphere detector is shown in figure
\ref{Fig_Setup}. Although the main results of the paper are valid
for any geometry of the dual system, we will consider the case of
two nested spheres with an inner radius $a$ and an external radius
$R$. The gap between the two bodies is taken small as compared to
these radii. The optomechanical sensor is based on a high-finesse
single-ended cavity, resonant with the incident laser beam. The
phase of the field reflected by the cavity is measured by a
homodyne detection.

As shown in \cite{Heidmann97,Pinard99} the motion of one sphere
(full or hollow one) induces a global phase-shift on the light
proportional to the displacement $\widehat{u}$ of the mirror
surface averaged over the beam profile,
\begin{equation}
\widehat{u}\left( t\right) =\left\langle u_{\bot }\left( t\right)
,v_{0}^{2}\right\rangle ,
\end{equation}
where the brackets stand for the overlap integral on the mirror
surface,
\begin{equation}
\left\langle f,g\right\rangle =\int d^{2}rf\left( r\right) g\left(
r\right) .
\end{equation}
$u_{\bot }\left( r,t\right) $ is the normal component of the
displacement at point $r$ and time $t$, and $v_{0}^{2}\left(
r\right) $ is the transverse gaussian structure of the light
intensity on the mirror,
\begin{equation}
v_{0}^{2}\left( r\right) =\frac{2}{\pi
w_{0}^{2}}e^{-2r^{2}/w_{0}^{2}}, \label{Equ_v0}
\end{equation}
where $w_{0}$ is the beam-spot size, identical for the two mirrors
in the case of a symmetric cavity.

The optomechanical sensor then reads out the differential
displacement $\widehat{u}_{o}\left( t\right)
-\widehat{u}_{i}\left( t\right) $ between the internal surface of
the outer sphere (subscript $o$) and the surface of the inner
sphere (subscript $i$).

A gravitational wave induces a displacement which can be written as a sum over all quadrupole modes \{n,2\} of the spheres \cite{Lobo95,Coccia98}. As long as the beam-spot size is small compared to the sphere radii, the spatial overlap between these modes and the beam profile is independent of $n$ and of the spot size. The radial displacement
of each sphere, for a gravitational wave of amplitude $\widetilde{h}\left[ \Omega \right] $ at frequency $\Omega $, is then equal to \cite{note},
\begin{equation}
\widehat{u}^{gw}\left[ \Omega\right]
=-\frac{1}{2}\sum_{n=1}^{\infty }b_{n}A_{n2}\left( a\right) \Omega
^{2}L_{n2}\left[ \Omega \right] \widetilde{h}\left[ \Omega \right]
, \label{Equ_ugw}
\end{equation}
where $b_{n}$ are the coefficients in the orthogonal expansion of
the response function of the sphere, $A_{n2}\left( a\right) $ are
the radial functions at the surface position $a$ (assumed to be the same for both surfaces since the gap is small), and
$L_{n2}\left[ \Omega \right]$ are the frequency dependences of the
modes \cite{Lobo95,Coccia98}. They correspond to harmonic
oscillators with resonance frequencies $\Omega _{n2}$ and loss
angle $\phi$, assumed to be the same for all modes and independent
of frequency,
\begin{equation}
L_{n2}\left[ \Omega \right] =\frac{1}{\Omega _{n2}^{2}-\Omega
^{2}-i\Omega _{n2}^{2}\phi }. \label{Equ_Ln2}
\end{equation}

To illustrate the relevant features of the detector response we
consider a dual-sphere made of Beryllium with an inner radius
$a=1.2$ m and an outer radius $R=2$ m. For these parameters the
fundamental frequencies are equal to $1161$ Hz for the outer
sphere and to $3075$ Hz for the inner one. Products
$b_{1}A_{12}\left( a\right) $ are equal to $1$ for the hollow
sphere and to $0.6$ for the solid one. As a consequence, the
displacements $\widehat{u}_{i}^{gw}$ and $\widehat{u}_{o}^{gw}$
are in phase for frequencies outside the two resonances and are
out of phase by $\pi $ radians in the intermediate frequency
domain between the two gravitational-waves sensitive modes. For
these frequencies, effects of both modes are added to each other
in the measured difference
$\widehat{u}_{o}^{gw}-\widehat{u}_{i}^{gw}$.

\begin{figure}
{\resizebox{6.5 cm}{!}{\includegraphics{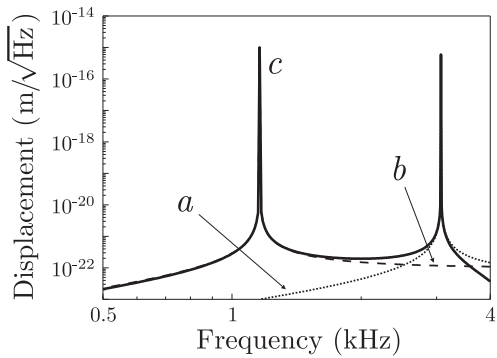}}}
\caption{\label{Fig_Signal}Displacement $\left|
\widehat{u}_{o}^{gw} -\widehat{u}_{i}^{gw}\right| $ as a function
of frequency, for a gravitational wave of amplitude
$\widetilde{h}=10^{-22}$ Hz$^{-1/2}$. Curves correspond to the
response of the fundamental mode of the inner sphere ({\it a}), of
the fundamental mode of the outer sphere ({\it b}), and of the two
modes together ({\it c}).}
\end{figure}

Figure \ref{Fig_Signal} shows the displacements induced by a
gravitational wave of amplitude $\widetilde{h}=10^{-22}$
Hz$^{-1/2}$, for each fundamental mode (curves {\it a} and {\it
b}). Curve ({\it c}) is the global response due to the two
fundamental modes. The response enhancement in the intermediate
frequency domain is clearly visible, resulting in a flat response
between the two fundamental resonances.

We have also determined the effect of higher modes by a numerical
computation of their mechanical characteristics \cite{Lobo95}.
Taking into account modes with resonance frequencies up to $30$
kHz for the solid sphere and to $16$ kHz for the hollow one, we
obtain no significant change for the response in the frequency
band between the two fundamental modes. This is related to the
frequency dependence $\Omega ^{2}/\Omega _{n2}^{2}$ in the
contribution of higher modes ($\Omega _{n2}\gg \Omega $) in the
sum (\ref{Equ_ugw}).

\section{Thermal and quantum noises}
\label{Noises}

We now determine the classical and quantum noises in the
measurement made by the optomechanical sensor. The phase of the
field reflected by the cavity reproduces the cavity length
variations. For a lossless cavity at resonance, the fluctuations
$\delta q^{out}\left[ \Omega \right] $ at frequency $\Omega $ for
the phase quadrature of the reflected field is given by
\cite{Heidmann97},
\begin{equation}
\delta q^{out}\left[ \Omega \right] =\delta q^{in}\left[ \Omega
\right] +16 \overline{\alpha
}^{in}\mathcal{F}\frac{\widehat{u}_{o}\left[ \Omega \right]
-\widehat{u}_{i}\left[ \Omega \right] }{\lambda },
\label{Equ_dqout}
\end{equation}
where $\delta q^{in}$ are the phase quadrature fluctuations of the
incident field, $\mathcal{F}$ the cavity finesse, $\lambda $ the
optical wavelength and $\overline{\alpha }^{in}$ the mean incident
field, related to the incident power $P^{in}$ and to the
wavevector $k=2\pi /\lambda$ by,
\begin{equation}
P^{in}=\hbar ck\left| \overline{\alpha }^{in}\right| ^{2}.
\end{equation}
We have neglected in eq. (\ref{Equ_dqout}) the low-pass filtering
due to the cavity bandwidth. Even for a cavity finesse
$\mathcal{F}=10^{6}$ and a gap between the two spheres of $1$ cm,
the cavity bandwidth is larger than $7.5$ kHz and has no
significant effect at frequencies of interest. We also assume that
the incident beam is in a coherent state, so that phase quadrature
$\delta q^{in}$ and intensity quadrature $\delta p^{in}$ are
uncorrelated with noise spectra given by \cite{Heidmann97},
\begin{equation}
S_{pp}^{in}\left[ \Omega \right] =S_{qq}^{in}\left[ \Omega \right]
= 1. \label{Equ_Sppin}
\end{equation}

The phase of the reflected beam provides a measurement of the
displacements induced by a gravitational wave, with additional
noises related to the incident phase fluctuations and to
displacement noises (eq. \ref{Equ_dqout}). The main sources of
displacement noise are the thermal fluctuations of spheres and the
back action of the measurement process due to the radiation
pressure exerted by the intracavity field on both mirrors. All
these noises are uncorrelated and the spectral strain sensitivity
is given by equating the contribution of the gravitational-wave
signal with the noise contributions,
\begin{equation}
S_{hh}\left[ \Omega \right] =\frac{S_{uu}^{shot}\left[ \Omega
\right] +S_{uu}^{th}\left[ \Omega \right] +S_{uu}^{ba}\left[
\Omega \right] }{\left| \widehat{u}_{o}^{gw}\left[ \Omega \right]
-\widehat{u}_{i}^{gw}\left[ \Omega \right] \right| ^{2}/\left|
\widetilde{h}\left[ \Omega \right] \right| ^{2}}, \label{Equ_Shh}
\end{equation}
where $S_{uu}^{shot}$, $S_{uu}^{th}$ and $S_{uu}^{ba}$ are the
displacement noise spectra due to incident phase fluctuations,
thermal noise, and back action, respectively.

We now determine these three noise spectra. The shot noise
$S_{uu}^{shot}$ is given by the equivalent displacement noise
corresponding in the measurement to the phase quadrature
fluctuations $\delta q^{in}$. From eqs. (\ref{Equ_dqout}) and
(\ref{Equ_Sppin}) one gets,
\begin{equation}
S_{uu}^{shot}\left[ \Omega \right] =\frac{\lambda
^{2}}{256\mathcal{F}^{2}\left| \overline{\alpha }^{in}\right|
^{2}}. \label{Equ_Suushot}
\end{equation}

Thermal fluctuations of the two spheres are uncorrelated and their
contributions are added to each other in the thermal noise
spectrum $S_{uu}^{th}$. For each sphere, the displacement can be
related to a Langevin force $F^{th}$ describing the coupling with
a thermal bath \cite{Landau58},
\begin{equation}
\widehat{u}^{th}\left[ \Omega \right] =\chi \left[ \Omega \right]
F^{th}\left[ \Omega \right] , \label{Equ_uth}
\end{equation}
where $\chi \left[ \Omega \right] $ is a mechanical susceptibility
characterizing the response of the sphere to an external force.
Since it relies on the displacement $\widehat{u}$ averaged over
the beam-spot size, it depends on the spatial overlap with the
light. Using a modal expansion for the displacements of the
sphere, $\chi \left[ \Omega \right] $ is given by
\cite{Heidmann97,Pinard99},
\begin{equation}
\chi \left[ \Omega \right] =\sum_{n=1}^{\infty}\left\langle
u_{n},v_{0}^{2}\right\rangle ^{2}\chi _{n}\left[ \Omega \right],
\label{Equ_Chi}
\end{equation}
where $\left\langle u_{n},v_{0}^{2}\right\rangle $ is the spatial
overlap between the mechanical mode $u_{n}(r)$ and the beam
profile (eq. \ref{Equ_v0}). $\chi _{n}\left[ \Omega \right] $ is
the mechanical susceptibility associated with mode $n$,
\begin{equation}
\chi _{n}\left[ \Omega \right] =\frac{1}{M_{n}\left( \Omega
_{n}^{2}-\Omega ^{2}-i\Omega _{n}^{2}\phi \right) },
\label{Equ_Chin}
\end{equation}
where $\Omega _{n}$ is the resonance frequency of the mode and
$M_{n}$ its effective mass related to the total mass $M$ and to
the volume $V$ of the sphere by,
\begin{equation}
M_{n}=\frac{M}{V}\int_{V}d^{3}r\left| u_{n}(r)\right| ^{2}.
\label{Equ_Mn}
\end{equation}

The noise spectrum of the Langevin force $F^{th}$ is related to
the mechanical susceptibility through the fluctuations-dissipation
theorem \cite{Landau58},
\begin{equation}
S_{FF}^{th}\left[ \Omega \right] =-\frac{2k_{B}T}{\Omega }{\rm
Im}\left( \chi \left[ \Omega \right] ^{-1}\right) ,
\label{Equ_SFFth}
\end{equation}
where $k_{B}$ is the Boltzmann constant and $T$ the temperature.
From eqs. (\ref{Equ_uth}) and (\ref{Equ_SFFth}) one then gets the
noise spectrum for the thermal fluctuations of the differential
displacement $\widehat{u}_{o}-\widehat{u}_{i}$ measured by the
optomechanical sensor,
\begin{equation}
S_{uu}^{th}\left[ \Omega \right] =\frac{2k_{B}T}{\Omega }{\rm
Im}\left( \chi _{o}\left[ \Omega \right] +\chi _{i}\left[ \Omega
\right] \right) , \label{Equ_Suuth}
\end{equation}
where $\chi _{o}$ and $\chi _{i}$ are the mechanical
susceptibilities of the outer and inner spheres, respectively.

Back-action effects are related to the radiation pressure forces
exerted by the intracavity field on both mirrors. Since the
radiation pressure has the same spatial profile as the intracavity
intensity, it can be shown \cite{Pinard99} that the displacement
$\widehat{u}^{ba}$ of each sphere is related to the force through
the same susceptibility as the thermal noise (eq. \ref{Equ_Chi}),
\begin{equation}
\widehat{u}^{ba}\left[ \Omega \right] =\chi \left[ \Omega \right]
F^{ba}\left[ \Omega \right] .
\end{equation}
The radiation pressure force is equal to,
\begin{equation}
F^{ba}\left[ \Omega \right] =2\hbar kI\left[ \Omega \right] ,
\end{equation}
where $I $ is the intracavity intensity normalized as a number of
photons per second. Intensity fluctuations can be related to the
amplitude quadrature of the incident field \cite{Heidmann97} and
one gets the noise spectrum of the radiation pressure,
\begin{equation}
S_{FF}^{ba}\left[ \Omega \right] =\frac{16\hbar ^{2}k^{2}}{\pi
^{2}}\mathcal{F}^{2}\left| \overline{\alpha }^{in}\right| ^{2}.
\label{Equ_SFFba}
\end{equation}

The radiation pressure forces exerted on the two mirrors are equal
and opposite. The induced cavity-length variation is therefore
related to the radiation pressure force through the sum of the
susceptibilities of the outer and inner resonators,
\begin{equation}
\widehat{u}_{o}^{ba}\left[ \Omega \right]
-\widehat{u}_{i}^{ba}\left[ \Omega \right] =\left( \chi _{o}\left[
\Omega \right] +\chi _{i}\left[ \Omega \right] \right)F^{ba}\left[
\Omega \right] ,
\end{equation}
and the resulting noise spectrum is given by,
\begin{equation}
S_{uu}^{ba}\left[ \Omega \right] =\left| \chi _{o}\left[ \Omega
\right] +\chi _{i}\left[ \Omega \right] \right|
^{2}S_{FF}^{ba}\left[ \Omega \right] . \label{Equ_Suuba}
\end{equation}

We have thus determined the spectra of the three fundamental
noises appearing in the measurement (eqs. \ref{Equ_Suushot},
\ref{Equ_Suuth} and \ref{Equ_Suuba}). They only depend on a few
parameters. The optical properties are accounted for via the
cavity finesse $\mathcal{F}$ and the incident light intensity
$\left| \overline{\alpha }^{in}\right| ^{2}$, with the usual
behavior that the shot noise is decreased for a larger cavity
finesse or a more intense light, whereas the back-action noise is
increased (eqs. \ref{Equ_Suushot} and \ref{Equ_SFFba}). This
behavior leads to the Standard Quantum Limit
\cite{Caves81,Braginsky92}. The mechanical characteristics only
depends on the sum $\chi _{o}+\chi _{i}$ of the susceptibilities
of the two spheres.

To understand the relevant features of the quantum noises we
consider a Beryllium dual-sphere at zero temperature
($S_{uu}^{th}=0$), neglecting all mechanical modes except the
fundamental modes of the two spheres. The sum in expression
(\ref{Equ_Chi}) of the mechanical susceptibilities is then limited
to the first term $n=1$, their frequency dependence being the same
as the response to a gravitational wave (compare eqs.
\ref{Equ_Ln2} and \ref{Equ_Chin}). In contrast to this response,
radiation pressure forces exerted on the two mirrors are opposite
and one gets a back-action noise cancellation in the intermediate
frequency domain between the two fundamental resonance
frequencies. The mechanical susceptibilities $\chi _{o}$ and $\chi
_{i}$ actually have opposite signs in between the two resonances
so that the effects of both spheres interfere destructively in the
global back-action noise $S_{uu}^{ba}$ (eq. \ref{Equ_Suuba}).

\begin{figure}
{\resizebox{6.5 cm}{!}{\includegraphics{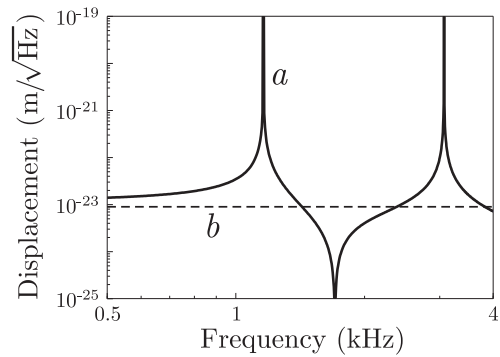}}}
\caption{\label{Fig_Quantum2}Displacement noises as a function of
frequency, for back-action effect ($\sqrt{S_{uu}^{ba}}$, curve
{\it a}) and for shot-noise ($\sqrt{S_{uu}^{shot}}$, curve {\it
b}). Only the two fundamental modes of the spheres are taken into
account.}
\end{figure}

Figure \ref{Fig_Quantum2} clearly shows this back-action
cancellation, resulting in an important noise reduction on the
whole frequency band between the two resonances (curve {\it a}).
The optical parameter $\mathcal{F}^{2}\left| \overline{\alpha
}^{in}\right| ^{2}$ is chosen in such a way that back-action and
shot noises are equal at low frequency. The frequency for which
exact cancellation occurs depends on the effective masses of the
fundamental modes (eqs. \ref{Equ_Chin} and \ref{Equ_Mn}) and can
be tuned in the intermediate frequency domain by changing the
mechanical characteristics of the spheres.

\begin{figure}
{\resizebox{6.5 cm}{!}{\includegraphics{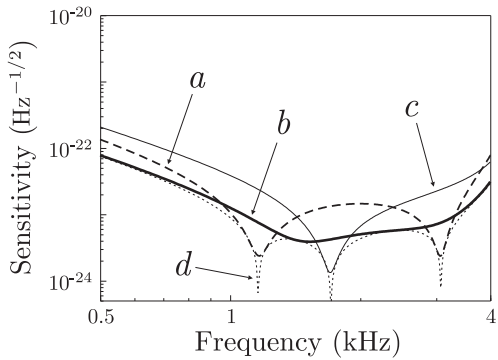}}}
\caption{\label{Fig_Sensit2}Quantum-limited spectral strain
sensitivity $\sqrt{S_{hh}}$ as a function of frequency, for
increasing incident intensities (curves {\it a} to {\it c}). Curve
{\it d} is the Standard Quantum Limit. Only the two fundamental
modes of the spheres are taken into account.}
\end{figure}

Since the contributions of both modes to the signal are added to
each other whereas the back-action noises are subtracted, one can
reach a very high sensitivity in the intermediate frequency
domain, as shown in figure \ref{Fig_Sensit2}. Curve ({\it b}) is
obtained for the same optical parameter as figure
\ref{Fig_Quantum2} and exhibits a high sensitivity over a wide
bandwidth. The behavior of the spectral strain sensitivity
actually depends on the ratio between the shot and back-action
noises. For a $10$-times smaller optical parameter (curve {\it
a}), shot noise is dominant and one recovers two peaks at the
fundamental resonances frequencies. For a $10$-times larger
optical parameter (curve {\it c}), the sensitivity follows the
back-action noise behavior and reaches an optimum value near the
frequency for which back-action is cancelled out.

Curve ({\it d}) in figure \ref{Fig_Sensit2} shows the Standard
Quantum Limit which corresponds to the minimum reachable noise for
a coherent incident light. It is obtained by adjusting at every
frequency the optical parameter $\mathcal{F}^{2}\left|
\overline{\alpha }^{in}\right| ^{2}$ in such a way that shot and
back-action noises are equal. From eqs. (\ref{Equ_Suushot}),
(\ref{Equ_SFFba}) and (\ref{Equ_Suuba}), the resulting minimum
displacement noise is given by,
\begin{equation}
S_{uu}^{sql}\left[ \Omega \right] =\hbar \left| \chi _{o}\left[
\Omega \right] +\chi _{i}\left[ \Omega \right] \right| .
\label{Equ_Suusql}
\end{equation}
As other noises, it depends on the mechanical characteristics only
via the sum of the susceptibilities of the two spheres.

\section{Mechanical susceptibility}
\label{Mechanical_Susceptibility}

To perform an accurate evaluation of the spectral strain
sensitivity of the detector, we need to calculate the mechanical
susceptibilities of the two spheres. For this purpose we can use
two different approaches, but both are insufficient for our needs.

The first one is the normal modes expansion
\cite{Bondu95,Gillespie95,Pinard99} which can be used
in principle in any case, if inhomogeneous losses are absent (it
fails in presence of inhomogeneous losses since correlations arise
between modes \cite{Majorana97,Conti02b,Yamamoto01}). This method
consists in computing the sum in eq. (\ref{Equ_Chi}) over a finite
number $N$ of modes,
\begin{equation}
\chi ^{\left( N\right)}\left[ \Omega \right]
=\sum_{n=1}^{N}\left\langle u_{n},v_{0}^{2}\right\rangle ^{2}\chi
_{n}\left[ \Omega \right]. \label{Equ_ChiSumN}
\end{equation}
This quantity of course converges towards the mechanical
susceptibility for a large number $N$ of modes. It unfortunately
converges very slowly, and the computation of the appropriate
overlap integrals between the modes and light profiles, for higher
and higher modes, is in practice too demanding.

The second approach applies in a direct manner the
fluctuations-dissipation theorem to estimate the thermal noise,
bypassing any normal mode expansion \cite{Levin98,Bondu98,Liu00}. This global approach has been worked out in the
limit of very low frequencies, so the results can be only used
well below any resonance frequency of the system. It actually
gives the static susceptibility in the case where the size of the
sensor is small compared to the dimensions of the system. For the
dual sphere, assuming the beam-spot size $w_{0}$ small compared to
the spheres radii, the sphere surface can be approximated by a
half-infinite plane and one gets \cite{Braginsky99},
\begin{equation}
\chi \left[ \Omega =0\right] =\frac{1-\sigma ^{2}}{\sqrt{\pi
}(1-i\phi )Ew_{0}}, \label{Equ_Chi0}
\end{equation}
where $E$ and $\sigma $ are respectively the Young modulus and the
Poisson coefficient of the sphere.

We want to calculate the susceptibilities for frequencies near or
between resonance frequencies of the system so that we cannot
apply this global approach. We instead consider a new method which
simultaneously uses the modal and global approaches. We write the
mechanical susceptibility in the form,
\begin{equation}
\chi \left[ \Omega \right] =\chi \left[ 0\right]
+\lim_{N\rightarrow \infty }\left( \chi ^{\left( N\right) }\left[
\Omega \right] -\chi ^{\left( N\right) }\left[ 0\right] \right) .
\label{Equ_ChiApprox}
\end{equation}
The first term is determined using the global approach (eq.
\ref{Equ_Chi0}) whereas the difference $\chi ^{\left( N\right)
}\left[ \Omega \right] -\chi ^{\left( N\right) }\left[ 0\right] $
is computed using the modal approach (eq. \ref{Equ_ChiSumN}). We
perform the sum over a finite number $N$ of modes such that the
resonance frequency $\Omega _{N}$ is much larger than the
frequency $\Omega $ of interest.

The advantage of this approach is that the difference in eq.
(\ref{Equ_ChiApprox}) converges more quickly than each term taken
separately. As a matter of fact, the difference is given by,
\begin{equation}
\chi ^{\left( N\right)}\left[ \Omega \right] -\chi ^{\left(
N\right)}\left[ 0 \right] =\sum_{n=1}^{N}\left\langle
u_{n},v_{0}^{2}\right\rangle ^{2}\left( \chi _{n}\left[ \Omega
\right] -\chi _{n}\left[ 0 \right] \right) . \label{Equ_ChiO0}
\end{equation}
From eq. (\ref{Equ_Chin}), the term in the sum for a mode $n$ such
that $\Omega _{n}\gg \Omega $ can be bounded by,
\begin{equation}
\chi _{n}\left[ \Omega \right] -\chi _{n}\left[ 0\right] \leq\chi
_{n}\left[ 0 \right] \frac{\Omega ^{2}}{\Omega _{n}^{2}-\Omega
^{2}} . \label{Equ_Bound1}
\end{equation}
The last factor $\Omega ^{2}/\left( \Omega _{n}^{2}-\Omega
^{2}\right) $ ensures a rapid convergence of the difference $\chi
^{\left( N\right) }\left[ \Omega \right] -\chi ^{\left( N\right)
}\left[ 0\right] $ as compared to the one of $\chi ^{\left(
N\right) }\left[ 0\right] $.

It is furthermore possible to determine an upper bound for the
error $\Delta \chi$ made by computing eq. (\ref{Equ_ChiO0}) up to
a mode $N$. $\Delta \chi$ is given by,
\begin{equation}
\Delta \chi \left[ \Omega \right] =\sum_{n=N+1}^{\infty
}\left\langle u_{n},v_{0}^{2}\right\rangle ^{2}\left( \chi
_{n}\left[ \Omega \right] -\chi _{n}\left[ 0 \right] \right) .
\end{equation}
Using eq. (\ref{Equ_Bound1}) for modes $n$ such that $\Omega _{n}\geq \Omega _{N} $ one gets,
\begin{equation}
\Delta \chi \left[ \Omega \right] \leq \chi \left[ 0\right]
\frac{\Omega ^{2}}{\Omega _{N}^{2}-\Omega ^{2}}.
\end{equation}
This upper limit shows that the computation is accurate within $1$ \% if the sum is extended up to modes with resonance frequencies $10$ times larger than the frequency $\Omega $ of interest.

The approach presented in this section can be used to determine
the thermal and back-action noises when one motion sensor is used
as read-out in any dual-resonator system. It can also be extended
to calculate the spatial correlations between modes. This would be
needed to discuss the effects on thermal and back-action noises,
when using a multi-transducer read-out configuration.

\section{Sensitivity of a dual sphere}
\label{Results}

We apply the approach presented in previous section to determine
the mechanical susceptibilities $\chi _{o}$ and $\chi _{i}$ for a
dual sphere. As previously we consider spheres made of Beryllium
with radii $a=1.2$ m and $R=2$ m. To calculate the mechanical susceptibilities we take into account the contributions of 120 modes for the solid sphere (resonance frequencies up to $30$ kHz) and of 80 modes for the hollow one (up to $16$ kHz) \cite{Lobo95}.

\begin{figure}
{\resizebox{6.5 cm}{!}{\includegraphics{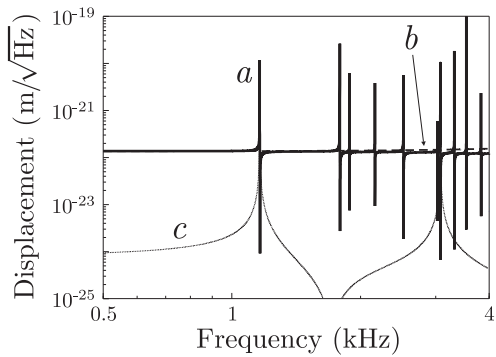}}}
\caption{\label{Fig_Quantum}Displacement noises as a function of
frequency, for back-action effect ($\sqrt{S_{uu}^{ba}}$, curve
{\it a}) and for shot-noise ($\sqrt{S_{uu}^{shot}}$, curve {\it
b}). Curve {\it c} shows the contribution of the two fundamental
modes to the back-action noise.}
\end{figure}

Figure \ref{Fig_Quantum} shows the shot and back-action noises
obtained by a numerical calculation of eq. (\ref{Equ_ChiApprox}).
Optical parameters are as follows: wavelength $\lambda =1$ $\mu$,
beam-spot size on the mirrors $w_{0}=1$ cm, cavity finesse
$\mathcal{F}=10^{6}$, and incident power $P^{in}=45$ mW. These
parameters are such that the back-action noise (curve {\it a}) and
the shot noise (curve {\it b}) are equal at low frequency.

Curve ({\it c}) in figure \ref{Fig_Quantum} shows the noise obtained when only the two fundamental modes of the spheres are considered, for the same optical parameters. Taking into account higher modes then drastically increases the back-action noise, by two orders of magnitude. The
resulting spectrum is almost flat in frequency and equal to the value
obtained at zero frequency by considering only the static
susceptibility $\chi \left[ 0\right] $ (eq. \ref{Equ_Chi0}). The
presence of resonant modes in the frequency band of interest leads
to narrow peaks in the noise spectrum. They correspond to the two
fundamental modes but also to other modes which are not sensitive
to gravitational waves. Back-action cancellation is responsible
for narrow dips after each resonance, at frequency where the
mechanical response of the resonant mode compensates the response
of all other modes.

\begin{figure}
{\resizebox{6.5 cm}{!}{\includegraphics{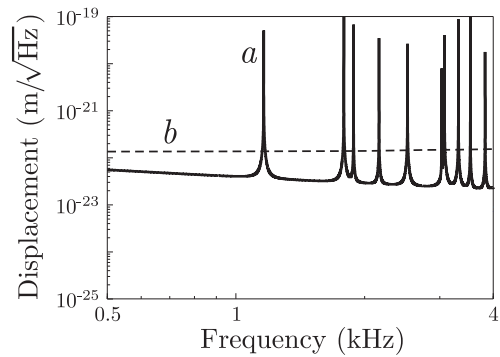}}}
\caption{\label{Fig_Thermal}Displacement noises as a function of
frequency, for thermal fluctuations ($\sqrt{S_{uu}^{th}}$, curve
{\it a}) and for shot-noise ($\sqrt{S_{uu}^{shot}}$, curve {\it
b}), in the same conditions as in figure \ref{Fig_Quantum}.}
\end{figure}

Figure \ref{Fig_Thermal} shows the thermal noise obtained for a
quality factor equal to $10^{7}$ ($\phi = 10^{-7}$) and a
temperature $T=0.1$ K. All modes of the dual sphere are thermally
excited and responsible for the presence of peaks in the spectrum.
For the chosen parameters the background thermal noise between two
resonances will have no significant effect since it is smaller
than the shot noise (curve {\it b} in figure \ref{Fig_Thermal}).

\begin{figure}
{\resizebox{6.5 cm}{!}{\includegraphics{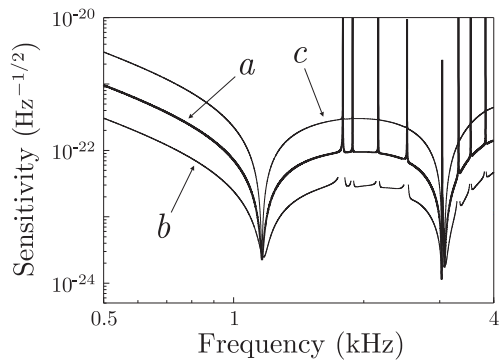}}}
\caption{\label{Fig_Sensit}Spectral strain sensitivity
$\sqrt{S_{hh}}$ as a function of frequency for the same parameters
as in figures \ref{Fig_Quantum} and \ref{Fig_Thermal} (curve {\it
a}). Curves {\it b} and {\it c} correspond to a beam-spot size of
$10$ cm and $0.1$ cm, instead of $1$ cm. Narrow peaks of these
curves have been removed for convenience.}
\end{figure}

Curve ({\it a}) of figure \ref{Fig_Sensit} shows the resulting
spectral strain sensitivity obtained from eq. (\ref{Equ_Shh}) by
computing the response to a gravitational wave as explained in section \ref{Optomechanical_Sensor}. Except for narrow peaks due to the resonant behavior of thermal and back-action noises, one gets a quantum-limited sensitivity with optimums of the order of $2\times 10^{-24}$ Hz$^{-1/2}$ at the resonance frequencies of the two fundamental modes. The
sensitivity is better than $10^{-22}$ Hz$^{-1/2}$ over a wide
frequency range, from $1$ kHz to $3.5$ kHz.

The frequency dependence of the sensitivity is however very
different from the one obtained in a two-modes analysis (figure
\ref{Fig_Sensit2}). This clearly shows that a two-modes or even a
few-modes analyses are not satisfactory to describe the mechanical
behavior of the spheres. Taking into account all the modes leads
to a larger mechanical response to radiation pressure
fluctuations. As compared to figure \ref{Fig_Sensit2} or to
previous results \cite{Cerdonio01} one gets sharper dips around
the fundamental resonance frequencies and a less flat sensitivity
between the two resonances.

Except near mechanical resonances, the response to quantum
fluctuations is very similar to the static response given by the
mechanical susceptibility $\chi \left[ 0\right]$. As shown in eq.
(\ref{Equ_Chi0}) a crucial parameter is then the beam-spot size
$w_{0}$ on the mirrors. Curves ({\it b}) and ({\it c}) in figure
\ref{Fig_Sensit} show the sensitivity for a beam-spot size of $10$
cm and $0.1$ cm, respectively. In both cases the optical
parameters are adjusted to stay at the standard quantum limit at
low frequency. Curve ({\it b}) corresponds to an incident power
$P^{in}=450$ mW and curve ({\it c}) to $4.5$ mW. It clearly
appears that increasing the beam-spot size by a factor $10$ leads
to wider dips at the fundamental resonance frequencies and to a
gain of sensitivity between the resonances by a factor
$\sqrt{10}$, as expected from the expression of the standard
quantum limit (eqs. \ref{Equ_Suusql} and \ref{Equ_Chi0}).

\section{Conclusion}
\label{Conclusion}

We have determined the sensitivity of a dual-sphere detector
taking into account the thermal, shot and back-action noises.
Neglecting all modes except the two fundamental ones, we obtained a very high sensitivity in the intermediate frequency domain between the two resonances, associated with a cancellation of back-action noise.

Using a new method mixing global and modal approaches, we have
calculated the mechanical response of the dual-sphere. The
features of the spectral strain sensitivity then look very
different from the idealized two-modes model, resulting in sharper
dips around the fundamental resonances and a less flat sensitivity
between the two resonances.

An important parameter is the beam-spot size $w_{0}$ on the
mirrors of the optomechanical sensor. Increasing $w_{0}$ reduces
the influence of high-frequency modes, resulting in a better
spectral strain sensitivity. A challenge is thus to design short
Fabry-Perot cavities with large beam waists. One may envision to
use a variant of conventional Fabry-Perot with large effective
waists \cite{Marin}. Another possibility is to use non-resonant
capacitive and inductive transducers at their quantum limits,
which intrinsically have large sensor area. Finally one may try to
implement a geometrically selective read-out, as proposed for
dual-cylinders \cite{Bonaldi02} to reduce the influence of
non-sensitive modes.


\begin{thebibliography}{0}
\bibitem{note} We assume that the gravitational wave impinges
with optimal direction and polarization with respect to the
position of the sensor. A spherical detector with a proper
multi-transducer read-out would be omnidirectional
\cite{Merkovitz97,Lobo00}
\end{thebibliography}
\end{document}